\documentclass{article} 
\usepackage{amsmath,amssymb,amsfonts}
\usepackage{epsfig}
\usepackage{nicefrac}
\usepackage{amsmath}
\usepackage{amssymb,amsfonts}
\usepackage{theorem}
\usepackage[latin1]{inputenc}
 
\begin{document}

\title{Superstring Theories}
\author{Constantin  Bachas and Jan Troost \\  \\ Laboratoire de Physique
 Th\'eorique \footnote{Unit\'e mixte
du CNRS et de l'Ecole Normale Sup\'erieure, UMR 8549.} \\
Ecole Normale Sup\'erieure \\ 24, rue Lhomond, 75005 Paris, France
}
\maketitle

\abstract{ This is  a short review of  superstring theories, highlighting the
important concepts,  developments   and  open problems of  the subject.  
}

 \section{Introduction}

String theory postulates  that all elementary particles in nature correspond to  different 
vibration states  of an underlying  relativistic string. 
In the quantum theory both the frequencies  and the amplitudes of
 vibration are quantized, so
that the quantum states of a string are discrete. They can be characterized by 
 their  mass, spin and  various gauge charges.   One of these  states
has  zero mass and spin equal to $2\hbar$,  and can 
be identified with the messenger  of gravitational interactions,  the   graviton. 
 Thus string theory is a candidate
for a {\bf unified theory}  of all fundamental interactions,  including {\bf quantum gravity}. 
 
 In this short review article  we discuss the theory of superstrings
 as consistent theories of quantum gravity. 
 The aim is to provide a quick (mostly  lexicographic and bibliographic) 
 entry to some of the salient features of the subject for a non-specialist
 audience.  Our treatment is thus neither complete nor comprehensive -- 
 there exist for this  several  excellent expert books, in particular by
 Green, Schwarz and Witten \cite{Green:1987sp}  
and by Polchinski \cite{Polchinski:1998rq}.  
An introductory textbook by Zwiebach \cite{Zwiebach:2004tj}
is also highly recommended for beginners.
Several other complementary  reviews on
 various aspects of superstring theories are available on the internet
  \cite{citedreviews}~; 
 some more will be given as we proceed.

\section{The five superstring theories}

 Theories of  relativistic extended objects are tightly constrained by { anomalies}, 
 i.e. quantum violations of classical symmetries. 
 These arise because the classical trajectory
of an extended  {\it p}-dimensional object
  (or  ``{\it p}-brane")  is described by the embedding  $X^\mu(\zeta^a)$, where
 $\zeta^{a=0,\cdots p}$ parametrize the brane worldvolume,  and $X^{\mu=0, \cdots D-1}$ are
 coordinates of the target space.   The quantum mechanics of a single
 {\it p}-brane is therefore
 a  ($p+1$)-dimensional quantum field theory,   and as such suffers a
 priori 
 from  ultraviolet
 divergences and anomalies.  The case $p=1$ is special in that these problems can be
 exactly handled.   The story for higher values of $p$ is much
 more complicated, as will become
apparent  later on. 
  
   The theory of  ordinary loops in space is  called closed {\bf bosonic string} theory. 
   The classical trajectory of a bosonic string  
    extremizes the {\bf Nambu-Goto} action (proportional to the invariant area of the worldsheet)
   \begin{equation} 
   S_{NG} =  - {1\over 2\pi\alpha^\prime} \int d^2\zeta 
   \; \sqrt{-{\rm det}(G_{\mu\nu}\; \partial_aX^\mu\partial_bX^\nu)}\ , 
  \end{equation}
  where $G_{\mu\nu}(X)$ is the target-space metric, and $\alpha^\prime$ is the Regge slope 
  (which is inversely proportional to the string tension and has
  dimensions
 of length squared). 
  In flat spacetime,   and for a conformal choice of worldsheet parameters
  $\zeta^\pm = \zeta^0\pm\zeta^1$,  the  equations of motion read:
  \begin{equation}\label{eq1}
  \partial_+\partial_- X^\mu = 0 \ \ \ {\rm and}\ \ \ 
  \eta_{\mu\nu}\partial_\pm X^\mu \partial_\pm X^\nu = 0\ ., 
  \end{equation}
 with $\eta_{\mu\nu}$ the Minkowski metric.
 The $X^\mu$ are thus free two-dimensional  fields,  subject to   
 quadratic  phase-space constraints known as the {\bf Virasoro  conditions}. 
 These can be solved consistently at the
 quantum level  in   the  {\bf critical dimension} $D=26$. Otherwise  the
 symmetries of eqs. (\ref{eq1}) are anomalous: either 
  Lorentz invariance is broken,  or  there is a  conformal 
  anomaly leading to unitarity problems.\footnote{
For  D $<$  26, unitary non-critical string theories
 in highly curved 
    rather than in the originally flat background can be constructed.}
   
  Even for $D=26$,  bosonic string theory is,  however,  sick  
 because its lowest-lying state  is a {\bf tachyon},  i.e. it has negative mass squared.  
 This follows from the zeroth-order Virasoro constraints, 
 \begin{equation}
 m^2 = - p^M p_M =   {4\over \alpha^\prime}\; (N_L -1) = 
 {4\over \alpha^\prime}\; (N_R -1) \ ,
 \end{equation}
 where $N_L$($N_R$)  is the sum of  the frequencies  of all left(right)-moving excitations
 on the string worldsheet. The negative contribution to $m^2$ 
  comes from quantum fluctuations, 
 and is analogous to  the  well-known Casimir energy. 
 The  tachyon  has $N_L=N_R=0$. Its presence signals an
 instability of Minkowski spacetime,    which in bosonic string theory 
 is expected to decay, possibly 
 to some lower-dimensional   highly-curved geometry.
 The details of how this happens are not, at present, well understood. 
 
 The  problem of the tachyon is  circumvented by endowing the string with additional,  
 anticommuting coordinates,  and requiring space-time {\bf supersymmetry} \cite{462}. 
 This is a symmetry that relates string states with integer spin,  obeying
 Bose-Einstein statistics,  to states with half-integer spin
 obeying Fermi-Dirac statistics.  There exist two standard
 descriptions of
 the superstring: the
 {\bf Ramond-Neveu-Schwarz} (RNS) formulation,  
 where the anticommuting coordinates $\psi^\mu$   carry a space-time  vector index, 
 and the {\bf Green-Schwarz} (GS) formulation in which they transform as a space-time spinor
 $\theta^\alpha$.  Each has  its advantages and drawbacks: the RNS formulation is
 simpler  from the  worldsheet point of view, but awkward for describing space-time 
 fermionic  states;  in the GS 
 formulation, on the other hand,  space-time supersymmetry
 is manifest  but  quantization can
  only be carried out in the restrictive  light-cone gauge.  A third  formulation,
   possibly combining
  the advantages of the other two, has been  proposed 
  more recently by Berkovits \cite{Berkovits:2002zk}
  -- it is at this stage  still being developed.

 Anomaly cancellation leads to
 five consistent superstring theories, all defined in $D=10$  flat space-time dimensions. 
 They are referred to as  type IIA, type IIB, heterotic $SO(32)$, heterotic  $E_8\times E_8$, 
 and type I.  The  two {\bf type II theories} are given (in the RNS formulation) 
  by a straightforward extension 
 of eqs. ({\ref{eq1})~:
  \begin{equation}\label{eq2}
  \partial_+\partial_- X^\mu = \partial_\mp \psi^\mu_\pm =  0 \ \ \ {\rm and}\ \ \ 
  \eta_{\mu\nu}\psi_\pm^\mu \partial_\pm X^\nu = 0\ . 
  \end{equation}
 The left- and right-moving  worldsheet
 fermions can be separately periodic or antiperiodic -- these
 are known as Ramond (R) and Neveu-Schwarz (NS) boundary conditions.  
 Ramond fermions have zero modes obeying
 a Dirac $\gamma$-matrix algebra, and which must thus be
 represented on spinor space. As a result
  out of the four possible boundary conditions for
   $\psi_+^\mu$ and $\psi_-^\mu$, namely 
  NS-NS, R-R, NS-R or R-NS, the first two give rise to string states
  that are space-time bosons,  while the other two give rise to states that are
  space-time fermions.    
  Consistency of the theory further  requires that one only
  keep  states of definite worldsheet fermion parities -- an operation  known
  as the  {\bf GSO} (for Gliozzi-Scherk-Olive)  {\bf projection}. This operation
   removes  the would-be tachyon, and 
  acts  as  a chirality projection  on the spinors. 
 The IIA and IIB theories differ only in the
  relative chiralities of  spinors coming from 
  the left  and right Ramond sectors. 
 
 The fact that string excitations split naturally into   non-interacting left  and
 right movers is crucial for  the construction of the {\bf heterotic strings}.  The key idea 
   is to put
together the left-moving sector  of the $D=10$ type II superstring and   the  right-moving
 sector of the $D=26$ bosonic string.  A subtlety arises because the left-right asymmetry
may lead to extra anomalies,   under global reparametrizations of the string worldsheet.  
 These are known as  {\bf modular anomalies}, and we will 
 come back  to them  in the following section.  Their
 cancellation  imposes stringent constraints on the zero modes of the unmatched (chiral)
 bosons in  the right-moving sector. The free-field expansion of these bosons 
 can be written as~: 
    \begin{equation}
 {\bf X}(\zeta^-)  =  {\bf x}_{\hskip -1.3 mm \ _{R}} + 
 \alpha^\prime {\bf p}_{\hskip -1.3 mm \ _{R}}\;   \zeta^- +  \sqrt{\alpha^\prime\over 2}
 \sum_{n\not= 0}  {i\over n}\; {\bf a}_n  \; e^{-2in\zeta^-}\ \   , 
  \end{equation}
where bold-face letters denote sixteen-component vectors. 
 Modular invariance then requires that the  generalized momentum 
 ${\bf p}_{\hskip -1.3 mm \ _{R}}$ 
 take its values in a sixteen-dimensional,  even self-dual  lattice. 
   There exist  two such lattices, and they are   generated by the 
 roots  of the Lie groups  ${\rm Spin}(32)/Z_2$ and $E_8\times E_8$.   They give rise to
 the two consistent heterotic string theories. 
  
      In contrast to the type II and heterotic theories,
 which are based on oriented closed strings, 
 the {\bf {type I  theory}}  has  unoriented closed
 strings as well as  open strings in its perturbative spectrum. 
 The closed strings are the same as in
 type IIB, except that one only keeps those states that are invariant under orientation reversal
  ($\zeta^+\leftrightarrow\zeta^-$).
 Open strings must also be invariant under this flip, and can furthermore
  carry  point-like ({\bf Chan-Paton}) 
  charges at their  two endpoints. This is analogous to the flavor
 carried
 by quarks at the endpoints
  of the chromoelectric flux tubes in QCD.  
  Ultraviolet finiteness requires that the { Chan-Paton charges}
 span a 32-dimensional vector space,
  so that  open strings transform  in bi-fundamental symmetric or
 antisymmetric representations of
  $SO(32)$.
  For a thorough review of type I string theory see reference  \cite{AS}.


  \section{Interactions and effective theories}

\begin{figure}
\begin{center}
\ \psfig{file=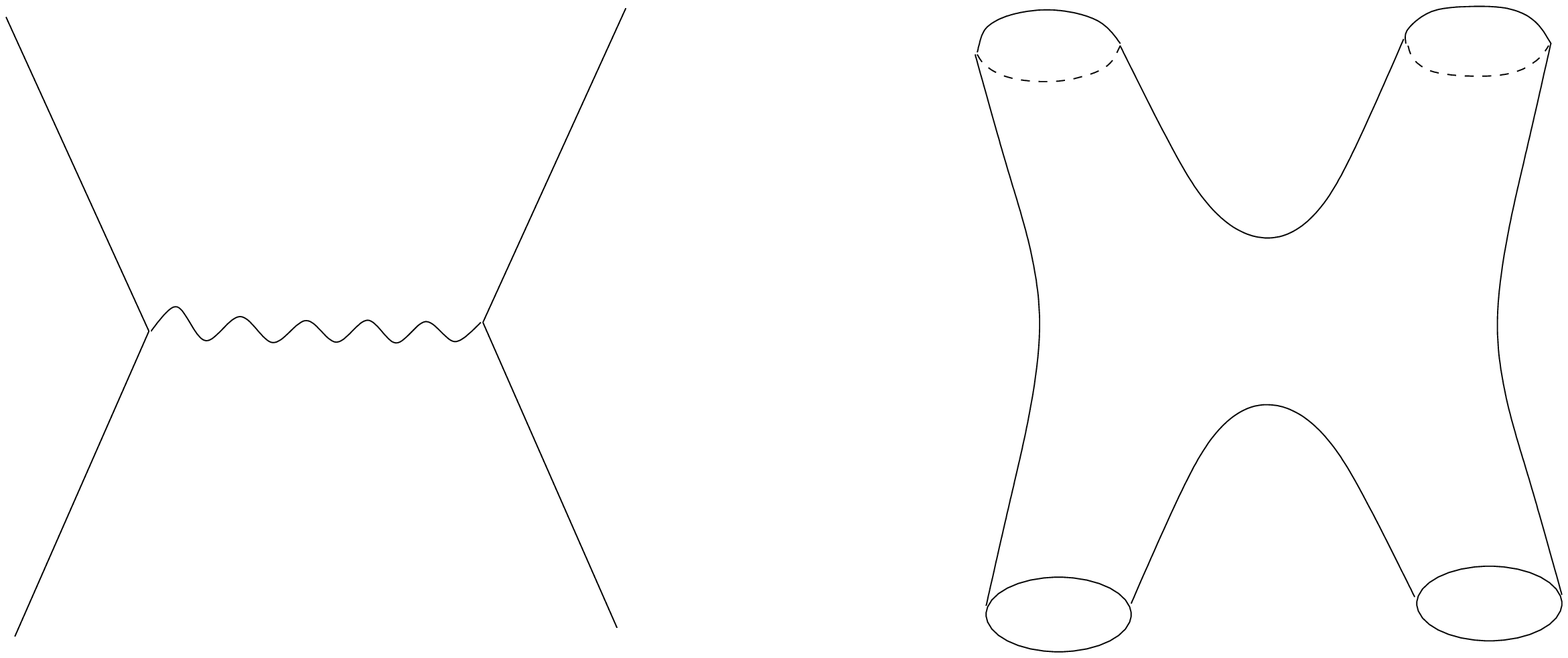, height=3.0cm}
\end{center}
\caption{\small A four-particle and a four-string interaction}
\label{figure:interaction}
\end{figure}
 
 Strings interact by splitting or by  joining at a point,  as is  illustrated in  figure 1.  
 This is a local  interaction  that respects the causality of the theory. 
 To compute scattering amplitudes one sums   over all worldsheets with a given
 set of asymptotic states, and weighs each local interaction with a factor of the {\bf string
 coupling constant}  $\lambda$.    The expansion
 in powers of $\lambda$  is analogous to the Feynman-diagram
 expansion of  point-particle  field theories. 
 These latter are usually defined by a Lagrangian, 
 or more exactly by a functional-integral measure, and  they make sense both for off-shell
 quantities as well as at the non-perturbative level. 
 In contrast, our current formulation of superstring
 theory is in terms of a perturbatively-defined S-matrix.
  The advent of dualities has offered glimpses
 of an underlying non-perturbative structure called   {\bf M-theory},
 but defining it precisely  is one of the major
 outstanding problems in the subject.\footnote{One approach
 consists in trying to define a second-quantized
 string field theory.   This is reviewed in the contribution  \cite{265} in the present volume}
 
       Another important expansion of string theory, very useful
 when it comes to extracting space-time properties, 
      is in terms of the characteristic string length 
 ${\it  l}_s = \sqrt{\alpha^\prime}$.  At energy scales
      $E {\it l}_s \ll 1$ only a handful of massless string
 states propagate, and their  interactions are
      governed by an effective low-energy Lagrangian.  In 
 the type-II theories the massless bosonic states (or
      rather their corresponding fields) consist of  the metric
  $G_{\mu\nu}$, a scalar field $\Phi$ called the 
       {\bf dilaton},  and a collection of {\bf antisymmetric $n$-form fields}
      coming from 
 both the NS-NS and the R-R sectors.  
    For type IIA these latter are a NS-NS  2-form $B_2$,   a  R-R 1-form $C_1$,  and a R-R 3-form
    $C_3$.  The leading-order action  for these fields reads~:
    \begin{eqnarray}\label{IIA}
    S_{\rm IIA} =&&\hskip -6mm
     {1\over 2\kappa^2}\int d^{10}x \Bigl[ \sqrt{-G}\;
 e^{-2\Phi} (R + 4\partial_\mu\Phi \partial^\mu\Phi
    - {1\over 2} \vert H_3\vert^2  )\cr
    &&  \hskip -3mm  -  \sqrt{-G}\;  ( {1\over 2}\vert F_2\vert^2 +
  {1\over 2} \vert F_4 - C_1\wedge H_3\vert^2) -  
    {1\over 2} B_2\wedge F_4\wedge F_4 \Bigr]  ,  
    \end{eqnarray}
    where $F_2= dC_1$, $H_3=dB_2$ and $F_4=dC_3$ are field strengths,  the wedge denotes the
     exterior product of forms,  
    and $\vert F_n\vert^2 = {1\over n!} F_{\mu_1\cdots \mu_n} F^{\mu_1\cdots \mu_n}$. 
    The dimensionful coupling $\kappa$ can be expressed in terms of the string-theory parameters,
    $2\kappa^2 =  (2\pi)^7 \lambda^2  {\alpha^\prime}^{4}$.
  A similar expression can be written for the IIB
    theory,  whose R-R sector contains a 0-form, a 2-form and a 4-form potential,
     the latter with self-dual field strength.

          The action (\ref{IIA}),  together with its fermionic part, 
 defines the maximally-supersymmetric non-chiral
          extension of Einstein's gravity in ten dimensions
  called type-IIA {\bf supergravity}  \cite{314}. 
          The dilaton and all  antisymmetric tensor fields 
   belong to  the   supermultiplet of  the
          graviton -- they provide together  the same number of (bosonic) states   as  
          a  ten-dimensional  non-chiral gravitino.  
          Supersymmetry fixes furthermore completely all 
 two-derivative terms of  the action, so that the theory
          defined by (\ref{IIA}) is (almost)
 unique.\footnote{There exists in fact a massive extension of IIA supergravity, which
          is the low-energy limit of  string theory with 
 a non-vanishing  R-R ten-form field strength. }
             It is,  therefore,  not surprising that it 
 should emerge   as  the low-energy limit of 
          the (non-chiral) superstring theory. This latter provides, 
       however, an  ultraviolet completion of an otherwise non-renormalizable theory, 
       a completion which is,  at least perturbatively,   finite and consistent. 

\begin{figure}
\begin{center}
\ \psfig{file=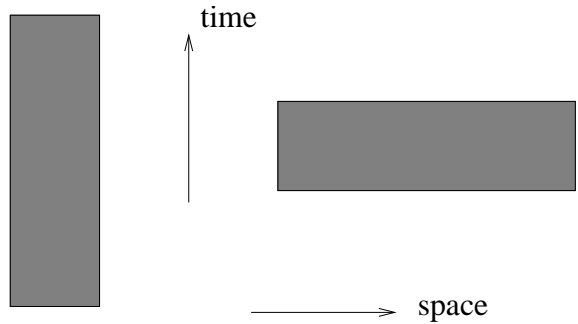, height=4.2cm}
\end{center}
\caption{\small The same torus diagram viewed in two different channels. }
\label{figure:torus}
\end{figure}

             The {\bf finiteness} of string perturbation
 theory has been, strictly-speaking, only  established up to two
             loops  -- for a recent review see  \cite{D'Hoker:2002gw}. 
  However, even though the technical problem
             is open and  hard, 
             the qualitative case  for all-order finiteness  is convincing.
               It can be  illustrated with the torus
             diagram which makes a one-loop contribution to  string amplitudes.   
             The thin torus  of figure 2 could  be traced either by a
 short, light string propagating (virtually) for
             a long time, or by a long,  heavy string propagating
 for a short period of time.  In conventional field
               theory these  two virtual trajectories would have made
               distinct contributions to the amplitude, one in the infrared
             and the second in the ultraviolet region. In string
 theory, on the other  hand,  they are  
              related by a  modular transformation 
 (that exchanges $\zeta^0$ with $\zeta^1$) and must not, 
              therefore,   be counted twice.
  A similar kind of argument shows that all potential divergences of
              string theory are infrared --
 they are therefore  kinematical (i.e. occur for special values of
the external momenta), or else they signal an instability of the
              vacuum and should  cancel if one expands around 
 a stable ground state.

       The low-energy limit of the 
heterotic and type I string theories is N=1 supergravity plus super
Yang-Mills. In addition to the N=1 graviton multiplet, the massless spectrum
now also includes gauge bosons and 
 their associated gauginos. The two-derivative effective  action in the
 heterotic case reads~: 
    \begin{eqnarray}\label{het}
    S_{\rm het} =&&\hskip -6mm
     {1\over 2\kappa^2}\int d^{10}x \sqrt{-G}\; e^{-2\Phi}
  \Bigl[ R + 4\partial_\mu\Phi \partial^\mu\Phi
   +  \;{\kappa^2\over g_{\rm YM}^2}  {\rm Tr} (
     F_{\mu\nu}F^{\mu\nu})   \cr
    &&  \hskip -1mm -  {1\over 2} \vert dB_2 - {\kappa^2\over g_{\rm YM}^2}\omega_3^{\rm
     gauge}   \vert^2\; 
  \Bigr] 
    + {\rm fermions}\     ,  
    \end{eqnarray}           
 where $\omega_3^{\rm gauge} = {\rm Tr}(A dA + {2\over 3} A^3)$ is the
 {\bf Chern-Simons}  gauge 3-form.
 Supersymmetry fixes again
 completely the above action --  the only freedom is in the choice of the
gauge group and of the Yang-Mills coupling  $g_{\rm YM}$. 
Thus, up to   redefinitions of the  fields,  the type
 I theory has necessarily the same low-energy limit.

      The D=10 supergravity plus super Yang-Mills has  a
      hexagon diagram that gives rise to gauge and gravitational 
      {\bf  anomalies},  similar to the triangle  anomaly in   D=4.
      It turns out that for the two special groups $E_8\times E_8$ and
      $SO(32)$,   the structure of these anomalies is  such that they
      can be cancelled by
      a combination of  local counterterms. One of them is of the form
 $ \int B_2\wedge X_8(F, R)$,  where $X_8$ is an 8-form quartic in  the curvature
      and/or  Yang-Mills field strength. The other is already present  in the
      lower 
line of expression (\ref{het}), with the replacement  $\omega_3^{\rm
      gauge} \rightarrow 
\omega_3^{\rm gauge}-\omega_3^{\rm Lorentz}$,  
   where  the second Chern-Simons form is built  out of the spin connection. 
Note that these modifications of the effective action 
      involve  terms with more  than two derivatives, and are not
      required by supersymmetry at the classical level.
 The discovery by  Green and Schwarz that string theory produces precisely
      these terms (from integrating out the massive string modes)
      was called the ``first superstring revolution.''


  \section{D-branes}
 
 A large window into the non-perturbative structure of string theory has been opened by
 the discovery of { D(irichlet)-branes},  and of strong/weak-coupling duality symmetries.  
 A D$p$-brane  is a  solitonic $p$-dimensional  excitation,  defined indirectly  
 by the property that its worldvolume   can
  attach open-string endpoints  (see figure 3).  Stable D$p$-branes exist in
  the type-IIA and type-IIB theories
  for $p$ even, respectively odd, and in the type I theory for $p=1$ and $5$. 
  They are charged under the R-R ($p+1$)-form potential or, for $p>4$, under  its magnetic dual. 
  Strictly-speaking,  only for $0\leq p\leq 6$ do D-branes resemble regular {\bf solitons}
  (the word stands for `solitary waves'). The
  D7- branes are more like cosmic strings, the
  D8-branes  are  domain walls, while the D9-branes
  are spacetime filling.    Indeed,  type-I string theory can be thought as arising from
   type-IIB   through  the introduction of an {\bf orientifold}  9-plane
   (required for tadpole cancellation)
   and of thirty two   D9-branes.

\begin{figure}
\begin{center}
\ \psfig{file=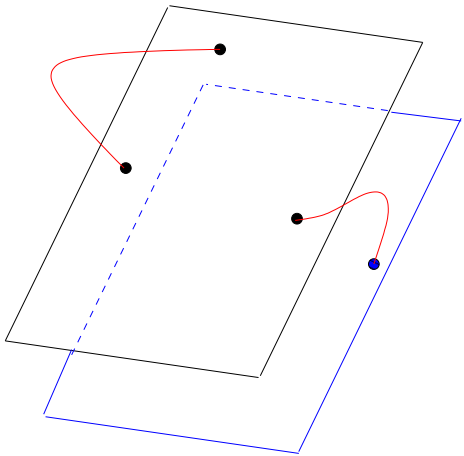, height=6cm}
\end{center}
\caption{\small D-branes and open strings}
\label{figure:dbranes}
\end{figure}

    The low-energy dynamics of a D$p$-brane is described by a supersymmetric abelian
     gauge  theory,   reduced from 10 down to $p+1$ dimensions. The gauge-field multiplet
     includes $9-p$ real scalars, plus gauginos in the spinor representation of the R-symmetry
     group $SO(9-p)$. These are precisely 
     the massless states of  an open string with  endpoints moving freely  on a hyperplane. 
     The   real scalar fields 
     are Goldstone modes of the broken translation invariance, i.e. they are 
    the transverse-coordinate fields $\vec Y(\xi^a)$ of the D-brane. 
     The bosonic part of the low-energy effective action 
 is the sum of a {\bf Dirac-Born-Infeld} (DBI) 
     and a Chern-Simons like  term~:
     \begin{equation}\label{Daction}
     I_{p} = - T_p  \int d^{p+1}\xi \; e^{-\Phi} 
  \sqrt{-{\rm det}( \hat G_{ab} + {\cal F}_{ab})} 
      -   \rho_p \int   \sum_n \hat C_n \wedge e^{\cal F}  \  , 
     \end{equation}
     where  ${\cal F}_{ab} = \hat B_{ab} + 2\pi\alpha^\prime F_{ab}$, 
     hats denote pullbacks on the brane of bulk tensor fields  (for example
     $\hat G_{ab} = G_{\mu\nu} \partial_a Y^\mu \partial_b Y^\nu$),
  $F_{ab}$ is the field strength
     of the worldvolume gauge field,  and in the CS  term one
     is instructed to keep    the $(p+1)$-form  of the expression under the integration sign. 
     The constants $T_p$ and $\rho_p$ are the tension and charge-density of the D-brane. 
 As was the case for the effective supergravities, the above
  action receives curvature corrections  that are  higher order
  in the $\alpha^\prime$ expansion.  Note however that a  class of higher-order terms
  have been already resummed in expression (\ref{Daction}). 
  These  involve arbitrary  powers of $F_{ab}$,  and  are closely related 
 (more precisely `{\bf T-dual}', see later)  to
  relativistic effects which  can be important even in the  weak-acceleration limit. 
  When refering
  to the D9-branes of the type I superstring, the action
  (\ref{Daction})  includes the Green-Schwarz terms required
  to cancel the gauge anomaly.

        The tension and charge-density of a D$p$-brane can be extracted from its coupling to the
(closed-string) graviton and   R-R ($p+1$)-form, with the result~:
\begin{equation}\label{tension}
T_p^2 = \rho_p^2 =  {\pi\over\kappa^2} (4\pi^2\alpha^\prime)^{3-p}\ . 
\end{equation}  
The equality of tension and charge follows from unbroken  supersymmetry,  and is also known as  
a Bogomol'nyi-Prasad-Sommerfield ({\bf BPS}) condition.   It implies that two or more 
identical D-branes exert 
no net static force on each other, because their  R-R repulsion cancels exactly
 their gravitational attraction.  A non-trivial check of the result (\ref{tension}) comes from
 the {\bf Dirac quantization}  condition (generalized to extended objects by Nepomechie
  and Teitelboim). Indeed, a 
  D$p$-brane and a D$(6-p)$-brane are dual excitations, like
 electric and magnetic charges in four dimensions, so their couplings  must obey
\begin{equation}
2\kappa^2  \rho_p \rho_{6-p} =  2\pi k\ \ \ {\rm  where}\ \ \ 
    k\in Z    \  . 
\end{equation}
 This ensures that the Dirac singularity of the long-range R-R fields of the branes
 does not lead to an observable
 Bohm-Aharonov phase.   The couplings  (\ref{tension})  obey 
this condition with $k=1$,  so  that D-branes  carry the smallest allowed R-R charges
  in  the  theory. 
  
   A simple but important  observation  is  that  open strings living 
   on a collection of $n$  identical D-branes   have   matrix-valued wavefunctions  $\psi_{ij}$, 
   where $i,j=1, \cdots, n$ label the possible  endpoints  of the string. The low-energy
   dynamics of the branes  is thus described by a non-abelian gauge theory, with group
   $U(n)$ if the open strings are oriented, and $SO(n)$ or $Sp(n)$ if they are not.  We have
   already encountered such  Chan-Paton factors in our discussion of the type I superstring.
    More generally, this simple property of D-branes 
      has lead to many   insights on 
   the {\bf geometric}  interpretation and {\bf engineering} of   gauge theories,  
    which are reviewed in the present  volume in  references
     \cite{243and360}.  It has also placed on a  firmer footing the idea of a 
  {\bf brane world},  according to which  the fields and interactions of the Standard Model 
  would be confined to a set  of D-branes, while gravitons are free to  propagate in the bulk 
   (for reviews see references \cite{13}\cite{133}). 
    It has, finally, inspired  the gauge/string-theory
   or {AdS/CFT}  correspondence \cite{245}\cite{2455} 
   on which we will comment  later on.

   
   \section{Dualities and M theory}

   One other key role  of D-branes has been to provide
evidence for the various {non-perturbative duality}
conjectures.   Dual descriptions of the same physics arise also in conventional field theory.
A prime example is the {\bf Montonen-Olive} duality of four-dimensional, N=4 supersymmetric
Yang-Mills, which is the low-energy theory describing the dynamics
of a collection of D3-branes.  The action
for the gauge field and six associated scalars $\Phi^I$ (all in the adjoint
representation of the gauge group $G$)  is
\begin{eqnarray}\label{N=4}
 S_{\rm N=4} = && \hskip -5mm
  - { 1\over 4 g^2}  \int d^4 x \;    {\rm Tr}  \Bigl (   F_{\mu\nu} F^{\mu\nu} 
 + 2 \sum_I D_\mu\Phi^I D^\mu\Phi^I  +  \sum_{I<J} 2  [\Phi^I, \Phi^J]^2\Bigr)  \cr
 &&\hskip -2mm
 - {\theta\over 32\pi^2}   \int d^4x  
 {\rm Tr}( F_{\mu\nu} \hskip -1mm \ ^*F^{\mu\nu})  +{\rm fermionic\ terms}
 \  . 
\end{eqnarray}
Consider for simplicity the case $G=SU(2)$. The scalar potential has
 flat directions 
along which  the
six $\Phi^I$ commute. By a $SO(6)$ R-symmetry rotation we can set all but one of them
 to zero, and let $<{\rm Tr}(\Phi^1\Phi^1)>= v^2$ in the vacuum. In
 this `Coulomb phase' of the theory 
 a $U(1)$ gauge multiplet stays massless, while the charged states become massive by the Higgs
effect.  The theory admits 
 furthermore smooth {\bf magnetic-monopole} and {\bf dyon}  solutions, 
and there is an elegant  formula for their mass~:
\begin{equation}
M = v \vert n_{\rm el} + \tau n_{\rm mg}\vert  , \ \ \ {\rm where}
\ \ \ \tau = {\theta\over 2\pi} + {4\pi i\over g^2}\   
\end{equation} 
and $n_{\rm el} (n_{\rm mg})$ denotes  the quantized electric (magnetic)  charge. 
This is a BPS  formula that receives no quantum corrections. It
exhibits the $SL(2,Z)$  covariance of the theory, 
\begin{equation}
\tau\to {a\tau+b\over c\tau+d}\ \  \ \ {\rm and} \ \  \ \ (n_{\rm el} , n_{\rm mg}) \to
(n_{\rm el} , n_{\rm mg})  \Bigl( \begin{matrix} a & b \cr c & d\end{matrix}
\Bigr)^{-1}\ .
\end{equation}
Here $a,b,c,d$ are integers subject to the condition $ad-bc=1$. 
 Of special importance is the   transformation $\tau\to -1/\tau$,  which
 exchanges electric and magnetic charges and (at least for
  $\theta =0$)  the strong-  with the weak-coupling regimes.  For more details 
 see the review \cite{harvey}. 
 
  The extension of these ideas to string theory can be illustrated with the strong/weak-
  coupling duality between the type I theory, and the  Spin$(32)/Z_2$
heterotic string. Both have the same massless spectrum and low-energy
action, whose form  is dictated entirely by
supersymmetry. The  only difference lies in the relations between  the 
string  and supergravity parameters. Eliminating the latter one finds~:
\begin{equation}
\lambda_{\rm het} = {1\over 2\lambda_{I}}\ \ \ {\rm and} \ \ \ 
\alpha^\prime_{\rm het} = \sqrt{2}\; \lambda_{I}\alpha^\prime_{I}\ ,  
\end{equation}
It is,  thus, very  tempting to conjecture that
the strongly-coupled type I theory has a dual description as a
weakly-coupled heterotic string. These are, indeed,  the only known
ulraviolet completions of the theory (\ref{het}). Furthermore,  for $\lambda_I\gg 1$ the
D1-brane  of the type I theory becomes light,  and could be 
plausibly identified with the
heterotic string. This  conjecture has been 
tested successfully by comparing various {\bf supersymmetry-protected}
quantities (such as the
tensions of BPS  excitations and special
higher-derivative terms in the effective action),  which 
can be calculated exactly either semiclassically, or
 at a given order in the perturbative expansion. Testing  the
duality for  non-protected quantities 
is a hard and important problem,  which looks  currently  out of reach. 

 The other three string theories have also well-motivated dual
 descriptions at strong coupling $\lambda$. 
   The  type IIB theory is believed to have a $SL(2,Z)$  symmetry,  similar
 to that of the N=4 super Yang Mills.\footnote{Note that $\lambda$ is
 a dynamical parameter, that changes with the vacuum expectation value
  of the dilaton $<\phi>$. Thus 
 dualities  are discrete gauge symmetries of  string theory.}  
The type IIA theory has a more surprising strong-coupling limit:
 it  grows  one  extra  dimension (of radius $R_{11} =
 1/\lambda \sqrt{\alpha^\prime}$), 
and can be  approximated at  low energy by  the maximal {\bf eleven-dimensional
 supergravity} of  Cremmer, Julia and Scherk. This latter  is a very
 economical  theory -- its massless bosonic fields are only the graviton and a 
three-form potential $A_3$. The bosonic part of the action reads~:
\begin{equation}
S_{\rm 11 D} = {1\over 2\kappa_{11}^2} \int d^{11}x
\sqrt{-G}\;  (R - {1\over 2}\vert F_4\vert^2) - {1\over 12\kappa_{11}^2}
\int A_3\wedge F_4\wedge F_4\ . 
\end{equation}
The  electric and magnetic charges of the three-form are a
(fundamental?) {\bf membrane} and a {\bf solitonic fivebrane}.
Standard Kaluza-Klein reduction on a circle \cite{duff}
maps $S_{\rm 11 D}$  to the IIA supergravity action  (\ref{IIA}), where
$G_{\mu\nu}$, $\phi$ and $C_1$ descend  from the eleven-dimensional  graviton, and
$B_2$ and $C_3$ from the three-form $A_3$.  Furthermore, all BPS excitations
of the type IIA string theory have a counterpart in eleven dimensions, as
summarized in the table below. 
Finally, if one compactifies the eleventh dimension on
an interval (rather than a circle),  one finds  the conjectured
strong-coupling limit of the $E_8\times E_8$ heterotic string.

\vskip 0.5cm 
\begin{table}[htp]  {\small
\begin{center}
\begin{tabular}{|c|c||c|c|}
 \hline
& & &  \\
{\bf tension}&  {\bf type-IIA}  & {\bf  ${\cal M}$  on $S^1$}
  &  {\bf tension } \\
&  & &  \\
\hline \hline
 & & & \\
$\displaystyle{\sqrt{\pi}\over  \kappa_{10}} 
\textstyle (2\pi\sqrt{\alpha^\prime})^3 $
 & D0-brane &  K-K excitation & $\displaystyle {1\over  R_{11}}$  \\
 & &  &\\
 \hline
 & &  & \\
$T_F= (2\pi\alpha^\prime)^{-1}$ & string  & wrapped membrane  &
$2\pi R_{11} \displaystyle \left( {2 \pi^2 \over \kappa_{11}^{\
      2}}\right)^{1/3} $ 
  \\
 & & & \\ 
 \hline
 & & & \\
$\displaystyle{\sqrt{\pi}\over  \kappa_{10}}\textstyle
 (2\pi\sqrt{\alpha^\prime})$ & D2-brane  & membrane  &
$ T_2^M= \displaystyle \left( {2 \pi^2 \over \kappa_{11}^{\
      2}}\right)^{1/3} $  \\
& &  &\\
  \hline
 & & & \\
$\displaystyle{\sqrt{\pi}\over  \kappa_{10}}\textstyle 
 (2\pi\sqrt{\alpha^\prime})^{-1}$
 & D4-brane   & wrapped five-brane &
$R_{11} \displaystyle \left( {2 \pi^2 \over \kappa_{11}^{\
      2}}\right)^{2/3} $
 \\ 
& & & \\ 
 \hline
 & & & \\
$\displaystyle {\pi \over \kappa_{10}^{\ 2}}\textstyle
 (2\pi\alpha^\prime)$ & NS-five-brane & five-brane  & 
$ \displaystyle {1\over 2\pi} \left( {2 \pi^2 \over \kappa_{11}^{\
      2}}\right)^{2/3} $
 \\
& &  &\\
  \hline
 & & & \\
$\displaystyle{\sqrt{\pi}\over  \kappa_{10}}\textstyle
 (2\pi\sqrt{\alpha^\prime})^{-3}$
 & D6-brane   & K-K monopole & 
$ \displaystyle  {2 \pi^2 R_{11}^{\ 2}  \over \kappa_{11}^{\  2}} $ \\ 
& & & \\ 
 \hline
\end{tabular}
\end{center} }
\vskip 0.4cm
\caption{\small  BPS excitations of type IIA string theory,  and their counterparts
in $\cal M$  theory compactified on a  circle of radius $R_{11}$.}
\end{table}

  The web of duality relations can be extended by compactifying further 
to $D\leq 9$ dimensions. Readers interested in more
details should consult Polchinski's book \cite{Polchinski:1998rq} or
one of the many existing reviews of the subject \cite{dual}\cite{citedreviews}.
In nine dimensions, in particular, the two type II  theories,  as well
as the two heterotic superstrings,  are pairwise T-dual. 
 {\bf T-duality} is a perturbative
symmetry (thus firmly established, not only conjectured)
which  exchanges {\bf momentum} and {\bf winding} modes. Putting
together all the links one arrives at the fully-connected web of
figure 4.
This makes the point that all  five consistent superstrings,  and also
eleven-dimensional   supergravity,   are  
limits of a unique  underlying structure 
called {\bf M theory}.\footnote{For lack of a better definition, 
`M'  is sometimes also used to denote the  D=11 supergravity plus supermembranes,
as in figure 4.}  
A background-independent
 definition of M-theory has remained  elusive. 
Attempts to define it as  a {\bf matrix model} of D0-branes, or by
quantizing a fundamental membrane, proved  interesting but incomplete. 
The difficulty stems from  the fact that in a generic background,
or in $D=11$ Minkowski spacetime, there is only a dimensionful
parameter 
fixing the scale at which the theory becomes strongly-coupled.

  \begin{figure}
\begin{center}
\ \psfig{file=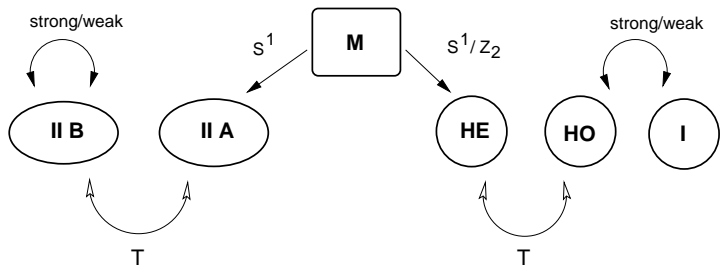, height=3.5cm}
\end{center}
\caption{\small Web of dualities in nine dimensions. }
\label{figure:dualities}
\end{figure}


 \section{Other developments and outlook}
  
We have not discussed in this brief review some important developments
covered in other contributions to the encyclopedia. 
For the reader's convenience,
and for completeness, we enumerate (some of)  them  giving
the  appropriate cross-references:

 \vskip 2mm   $\bullet$ {\bf Compactification}. To make contact with the Standard
   Model of particle physics, one has to compactify string theory on a
   six-dimensional manifold. There is an embarassement of riches, but  no
   completely realistic  vacuum and, more significantly,   
   no guiding dynamical principle to help us decide -- see \cite{douglas}. 
   The controlled (and phenomenologically required) breaking of spacetime
   supersymmetry is also a problem.

   \vskip 2mm $\bullet$ {\bf Conformal field theory and quantum
   geometry}. The algebraic tools of 2D conformal field theory, both bulk and boundary
   -- see \cite{conf},  play an important role in string theory. They
   allow,  in certain cases, a resummation of $\alpha^\prime$ effects,  
   thereby probing  the   regime where classical geometric notions do not apply.

   \vskip 2mm $\bullet$ {\bf Microscopic models of black holes}. Charged 
   extremal black holes can be  modeled in string theory
  by  BPS configurations of D-branes. This has lead to the
     first microscopic derivation of the Bekenstein-Hawking entropy formula,  a result
     expected from any  consistent theory of quantum gravity.   
  As with the tests of duality, the
   extension of these results to neutral black holes  is a difficult open problem
    --  see \cite{bhs}.
 
  \vskip 2mm  $\bullet$ {\bf AdS/CFT and holography}. A new  type
   of (holographic)  duality is the one that relates  supersymmetric gauge 
   theories in four dimensions to string theory in asymptotically anti-de Sitter 
   spacetimes. The sharpest and best tested version of this duality relates
    N=4 super Yang Mills to string theory in $AdS_5\times S_5$.  Solving the $\sigma$-model
    in this latter background is one of the keys to further progress in the  subject 
   -- see \cite{245}.

   \vskip 2mm $\bullet$ {\bf String phenomenology}. Finding an experimental 
   confirmation of string theory
   is clearly  one of the most pressing outstanding questions. 
   There exist several interesting possibilities for this  --
   cosmic strings, large extra dimensions, modifications of gravity, primordial
   cosmology  -- see the review \cite{uranga}.
   Here we  point out the one supporting piece of experimental evidence~: the unification
   of the gauge couplings of the (supersymmetric, minimal)
   Standard Model at a scale close to, but below the Planck scale, as illustrated 
   in  figure  \ref{figure:couplingunification}. 
   This is a generic `prediction'  of  string theory, especially in its heterotic version. 

\vskip 5mm 

\begin{figure}
\begin{center}
\ \psfig{file=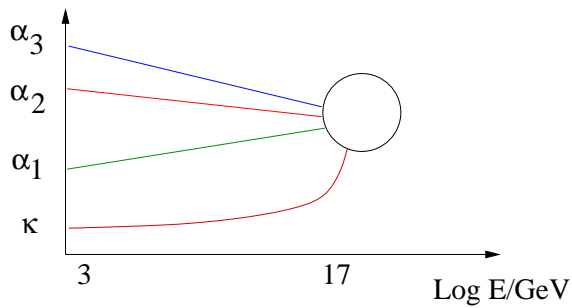, height=4cm}
\end{center}
\caption{\small The unification of couplings}
\label{figure:couplingunification}
\end{figure}

\end{document}